\documentclass[preprint,preprintnumbers,amsmath,amssymb]{revtex4}
\usepackage[dvipdfmx]{graphicx}
\usepackage{mathptmx,bm}
\usepackage[compatibility=false]{caption}

\begin{document}


\title{Spin waves and spin-state transitions in a ruthenate high-temperature antiferromagnet}

\author{H. Suzuki$^{1*}$, H. Gretarsson$^{1,2}$, H. Ishikawa$^{1,3}$, K. Ueda$^{1}$, Z. Yang$^{1}$, H. Liu$^{1}$,\\ H. Kim$^{1,4,5}$, D. Kukusta$^{1}$, A. Yaresko$^{1}$, M. Minola$^{1}$, J. A. Sears$^{2}$, \\S. Francoual$^{2}$, H.-C. Wille$^{2}$, J. Nuss$^{1}$, H. Takagi$^{1,3,6}$,  B. J. Kim$^{1,4,5}$,\\ G. Khaliullin$^{1}$, H. Yava\c{s}$^{2}$\footnote[3]{Current address: SLAC National Accelerator Laboratory, 2575 Sand Hill Rd, Menlo Park, CA 94025, USA} and B. Keimer$^{1*}$}
 
\affiliation{$^1$Max-Planck-Institut f\"{u}r Festk\"{o}rperforschung, Heisenbergstra\ss e 1, D-70569 Stuttgart, Germany}

\affiliation{$^2$Deutsches Elektronen-Synchrotron DESY, Notkestra\ss e 85, D-22607 Hamburg, Germany}

\affiliation{$^3$Institut f\"{u}r Funktionelle Materie und Quantentechnologien, Universit\"{a}t Stuttgart, 70569 Stuttgart, Germany}

\affiliation{$^4$Department of Physics, Pohang University of Science and Technology, Pohang 790-784, South Korea}

\affiliation{$^5$Center for Artificial Low Dimensional Electronic Systems, Institute for Basic Science (IBS), 77 Cheongam-Ro, Pohang 790-784, South Korea}

\affiliation{$^6$Department of Physics, University of Tokyo, Bunkyo-ku, Tokyo 113-0033, Japan}

\date{\today}

\maketitle

{\bf
Ruthenium compounds play prominent roles in materials research ranging from oxide electronics \cite{Koster.G_etal.Rev.-Mod.-Phys.2012} to catalysis \cite{Over.H_etal.Chem.-Rev.2012}, and serve as a platform for fundamental concepts such as spin-triplet superconductivity \cite{Maeno.Y_etal.J.-Phys.-Soc.-Jpn.2012}, Kitaev spin-liquids \cite{Jackeli.G_etal.Phys.-Rev.-Lett.2009,Chaloupka.J_etal.Phys.-Rev.-Lett.2010,Sandilands.L_etal.Phys.-Rev.-Lett.2015,Banerjee.A_etal.Nat.-Mater.2016}, and solid-state analogues of the Higgs mode in particle physics \cite{Jain.A_etal.Nat.-Phys.2017,Souliou.S_etal.Phys.-Rev.-Lett.2017}. However, basic questions about the electronic structure of ruthenates remain unanswered, because several key parameters (including the Hund's-rule, spin-orbit, and exchange interactions) are comparable in magnitude, and their interplay is poorly understood - partly due to difficulties in synthesizing sizable single crystals for spectroscopic experiments. Here we introduce a resonant inelastic x-ray scattering (RIXS) \cite{Kotani.A_etal.Rev.-Mod.-Phys.2001,Ament.L_etal.Rev.-Mod.-Phys.2011} technique capable of probing collective modes in microcrystals of $4d$-electron materials. We present a comprehensive set of data on spin waves and spin-state transitions in the honeycomb antiferromagnet SrRu$_{2}$O$_{6}$ \cite{Hiley.C_etal.Angew.-Chem.-Int.-Ed.2014}, which possesses an unusually high N\'eel temperature \cite{Tian.W_etal.Phys.-Rev.-B2015,Hiley.C_etal.Phys.-Rev.-B2015,Streltsov.S_etal.Phys.-Rev.-B2015,Singh.D_etal.Phys.-Rev.-B2015,Hariki.A_etal.Phys.-Rev.-B2017,Okamoto.S_etal.Sci.-Rep.2017}. The new RIXS method provides fresh insight into the unconventional magnetism of SrRu$_{2}$O$_{6}$, and enables momentum-resolved spectroscopy of a large class of $4d$ transition-metal compounds. 
}


\newcounter{Fig1}
\setcounter{Fig1}{1}
\newcounter{Fig2}
\setcounter{Fig2}{2}
\newcounter{Fig3}
\setcounter{Fig3}{3}
\newcounter{Fig4}
\setcounter{Fig4}{4}

Inelastic neutron scattering (INS) experiments on magnetic collective modes yield highly specific information on the energy scale and spatial range of the electronic correlations that drive magnetic order, spin-liquid behavior, and unconventional superconductivity in quantum materials. However, since cm$^{3}$-sized single crystals required for INS are difficult to grow for ruthenates and other $4d$ transition metal compounds, our current understanding of electronic exchange and correlation effects in these materials is quite limited. Recent advances in x-ray instrumentation have enabled RIXS experiments on dispersive magnetic excitations in $3d$ and $5d$ metal compounds with energy resolution approaching the one of INS, and sensitivity sufficient to probe microcrystals and thin-film structures \cite{Braicovich.L_etal.Phys.-Rev.-Lett.2009,Kim.J_etal.Phys.-Rev.-Lett.2012}. In these experiments, the incoming photons are tuned to the dipole-active $L$-absorption edges of the metal atoms in the soft (hard) x-ray regimes for $3d$ ($5d$) metals, and the energy distribution of the scattered x-rays is measured by high-resolution analyzers. Analogous experiments on $4d$ metal compounds have thus far not been possible, because their $L$-edges are in a  photon energy range between the soft and hard x-ray regimes (1-5 keV) that is difficult to access, and because suitable analyzers were not available.

We have built an intermediate-energy RIXS spectrometer (IRIXS) with an analyzer designed for measurements at the $L$-adsorption edges of $4d$ metal compounds. Using this instrument, we obtained momentum-resolved spectra of spin waves and spin-state transitions from a microcrystal of SrRu$_{2}$O$_{6}$, a compound based on ruthenium-oxide honeycomb layers \cite{Hiley.C_etal.Angew.-Chem.-Int.-Ed.2014} whose extraordinarily high N\'eel temperature of 563 K \cite{Tian.W_etal.Phys.-Rev.-B2015,Hiley.C_etal.Phys.-Rev.-B2015} has recently inspired considerable theoretical work \cite{Streltsov.S_etal.Phys.-Rev.-B2015,Singh.D_etal.Phys.-Rev.-B2015,Hariki.A_etal.Phys.-Rev.-B2017,Okamoto.S_etal.Sci.-Rep.2017}. Models proposed to explain this observation include extended molecular orbitals on the hexagons of the honeycomb lattice \cite{Streltsov.S_etal.Phys.-Rev.-B2015}, and a dichotomy of localized and itinerant valence electrons residing in different Ru orbitals \cite{Okamoto.S_etal.Sci.-Rep.2017}. We show that the spin-wave dispersions extracted from the IRIXS data are well described by a Heisenberg model with nearest-neighbor exchange interactions, suggesting instead that the valence electrons are in the localized regime. At higher excitation energies, we detect spin-state transitions that allow us to determine the Hund's rule interaction of the $4d$ valence electrons. These results establish IRIXS as an incisive probe of the low-energy electronic structure of a large variety of ruthenates and other $4d$ metal compounds.


Figure \arabic{Fig1}a illustrates the crystal structure of SrRu$_{2}$O$_{6}$, which belongs to the hexagonal space group $P\bar{3}1m$ with lattice constants $a$ = 5.200(2) \AA\ and $c$ = 5.225(2) \AA. The ruthenium ions are octahedrally coordinated by oxygen ions and form a two-dimensional honeycomb network. SrRu$_{2}$O$_{6}$ is insulating and exhibits G-type antiferromagnetism below $T_{N} = 563$ K \cite{Tian.W_etal.Phys.-Rev.-B2015,Hiley.C_etal.Phys.-Rev.-B2015}. The Ru spins are oriented along the $c$-axis (Fig. \arabic{Fig1}b). We have carried out complementary resonant X-ray diffraction measurements to confirm the long-range magnetic order in our samples (see Supplementary Information).

While the nominal Ru valence of 5+ in the octahedral crystal field dictates three electrons in the $4d$ $t_{2g}$ orbitals, a neutron powder diffraction study revealed a local magnetic moment of 1.4 $\mu_{B}$ per Ru atom \cite{Hiley.C_etal.Phys.-Rev.-B2015}, which is significantly lower than 3 $\mu_{B}$ expected for the high-spin $S=3/2$ configuration. We synthesized hexagonal single crystals with typical diameter of $\sim$ 50 $\mu$m (Fig. \arabic{Fig1}c) using a hydrothermal technique (see Methods), which cannot be scaled up to yield crystals with dimensions required for INS. Alternative spectroscopic tools are therefore required to map out the magnetic collective modes and test the diverse models that have been proposed to explain the high N\'eel temperature and the moment reduction in SrRu$_{2}$O$_{6}$ \cite{Streltsov.S_etal.Phys.-Rev.-B2015,Singh.D_etal.Phys.-Rev.-B2015,Hariki.A_etal.Phys.-Rev.-B2017,Okamoto.S_etal.Sci.-Rep.2017}. This situation (which is also encountered in numerous other ruthenium compounds) has motivated the development of the IRIXS spectrometer.



Figure \arabic{Fig1}d,e shows a schematic of the IRIXS spectrometer and the scattering geometry. The incident x-rays were monochromatized by a pair of asymmetric Si(111) crystals and focused onto the sample (see Methods for details). They are linearly $\pi$-polarized and the scattering angle $2\theta$ is fixed at 90$^{\circ}$. This geometry significantly suppresses the elastic charge (Thomson) scattering, allowing us to detect low-energy magnetic excitations. Since the Ru$_{2}$O$_{6}$ honeycomb layers are sandwiched by the chemically inactive Sr layers (Fig. \arabic{Fig1}b), the inter-plane magnetic interaction can be treated as weak, as we will see below. Hereafter, the momentum transfer is expressed in terms of the in-plane component, which was scanned by rotating the sample angle $\theta$. The measurement paths in the in-plane ${\bm q}$ space are shown in Fig. \arabic{Fig1}f. The azimuthal angle $\phi$ is fixed at 0$^{\circ}$ for the measurement along the (-$H$, 0) direction, and at -30$^{\circ}$ for the (-$H$, -$H$) direction.

\begin{figure}[htbp]
   \centering 
   \includegraphics[width=14cm]{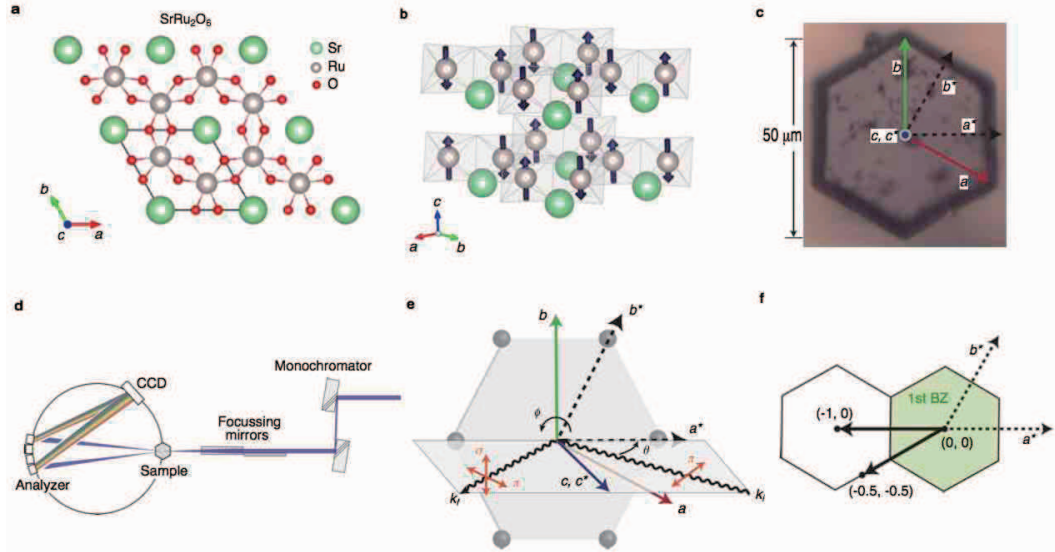}
   \caption*{\raggedright\textbf{Figure 1 $\vert$ Magnetic structure, IRIXS instrument, and scattering geometry}. {\bf a}, Crystal structure and {\bf b}, G-type antiferromagnetic order of SrRu$_{2}$O$_{6}$. The magnetic easy axis is along the crystallographic $c$-axis. {\bf c}, Photograph of a SrRu$_{2}$O$_{6}$ single crystal. {\bf d}, Schematic of the IRIXS spectrometer. {\bf e}, Scattering geometry for IRIXS experiments. The incoming x-rays are linearly $\pi$-polarized and the polarization of the outgoing x-rays is not analyzed. The scattering angle $2\theta$ is fixed at 90$^{\circ}$. {\bf f}, Measurement paths in the in-plane momentum space.}

  \label{Fig1}
 \end{figure}

Figure \arabic{Fig2}a,b shows raw (uncorrected) Ru $L_{3}$ RIXS spectra of SrRu$_{2}$O$_{6}$ along the (-$H$, 0) and (-$H$, -$H$) directions, respectively.  In contrast to soft x-ray RIXS, the large incident photon momentum of IRIXS allowed us to map out the dispersion relations of the electronic excitations over the entire first Brillouin zone.
The spectra exhibit three distinct features. First,  low-energy sinusoidally dispersive peaks (red circles in Fig. \arabic{Fig2}a,b) emanate from the antiferromagnetic ordering vector ${\bm q}=$ (-1, 0) and have maxima at ${\bm q}=$ (-0.5, 0) along the (-$H$, 0) direction and at (-0.33, -0.33) along the (-$H$, -$H$) direction. 
Second, there are additional pronounced peaks at $\sim$ 0.65 eV (green triangles in Fig. \arabic{Fig2}a,b), which do not show any dispersion within the experimental error.  Finally, one also observes a broad continuum with an onset of $\sim$ 0.4 eV. To visualize the momentum dependence of the scattering intensity, we show a color map of the RIXS data in Fig. \arabic{Fig2}c. The intensity of the low-energy branch is maximal at the antiferromagnetic ordering vector and vanishes at ${\bm q}=0$, as expected for spin waves (magnons). Based on the dispersion relation and the intensity distribution, the dispersive low-energy excitations can therefore be assigned to single magnons. As discussed below, we assign the nondispersive feature at higher energies to intra-ionic spin-state transitions between $S=3/2$ and $S=1/2$ configurations, and the continuum to electron-hole excitations across the Mott gap. Fig. \arabic{Fig2}d schematically summarizes the decomposition of the full IRIXS spectra.

\begin{figure}[htbp]
   \centering 
   \includegraphics[width=14cm]{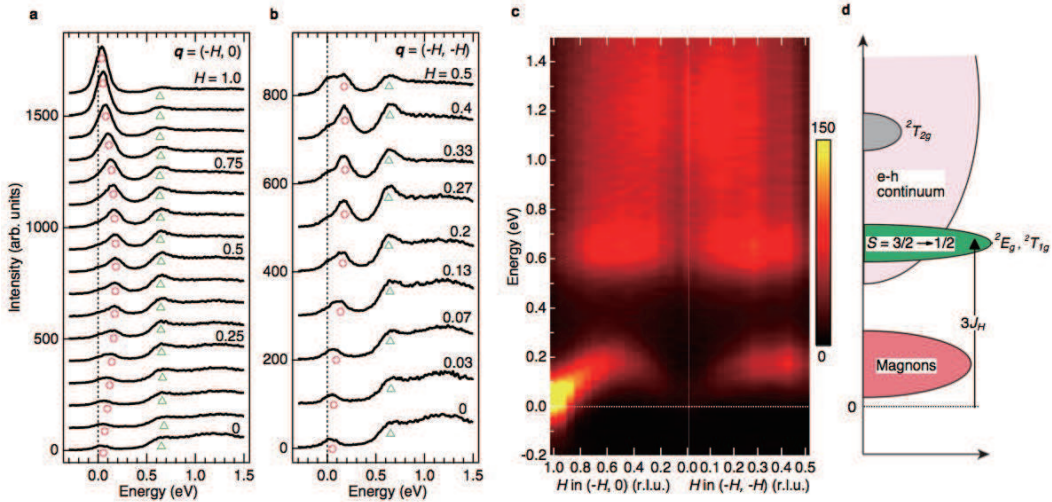}
   \caption*{\raggedright\textbf{Figure 2 $\vert$ IRIXS spectra of SrRu$_{2}$O$_{6}$}. {\bf a}, Raw Ru $L_{3}$ IRIXS spectra of SrRu$_{2}$O$_{6}$ along the (-$H, 0$) direction taken at 10 K, well below $T_{N}$. The in-plane momenta are equally spaced between ${\bm q}=$ (0, 0) and (-1, 0). The magnon peak positions determined by spectral fitting (Fig. \arabic{Fig3}) are denoted by red circles. The peak positions of the spin-state transitions, determined by finding the local maxima of smoothed spectra, are denoted by green triangles. {\bf b}, IRIXS spectra along the (-$H$, -$H$) direction.  {\bf c}, Color map of the IRIXS intensity in panels {\bf a} and {\bf b}. Elastic peaks were subtracted for better visualization of the low-energy features. {\bf d}, Schematic illustration of the elementary excitations.}
  \label{Fig2}
 \end{figure}



Having demonstrated the capability of IRIXS to detect single-magnon excitations, we now analyze the low-energy IRIXS spectra and discuss their implications for the magnetic interactions. Figure \arabic{Fig3}a,b shows the decomposition of the spectra into an elastic background (dotted black line) and the magnon peak (red line), based on least-squares fits to two Lorentzian profiles convoluted with the experimental resolution. The excellent agreement of the experimental data and the total intensity resulting from the fits (green lines) allows us to accurately determine the dispersion and spectral weight of the magnon feature as a function of ${\bm q}$ (Fig. \arabic{Fig3}c,d).  One notices a magnon gap of 36 meV at ${\bm q}=($-$1, 0)$, reflecting the $c$-axis spin anisotropy. The magnon branch disperses up to 183 meV, both at ${\bm q}=($-$0.5,$-$0.5)$ and $($-$0.33,$-$0.33)$.

\begin{figure}[htbp]
   \centering 
   \includegraphics[width=14cm]{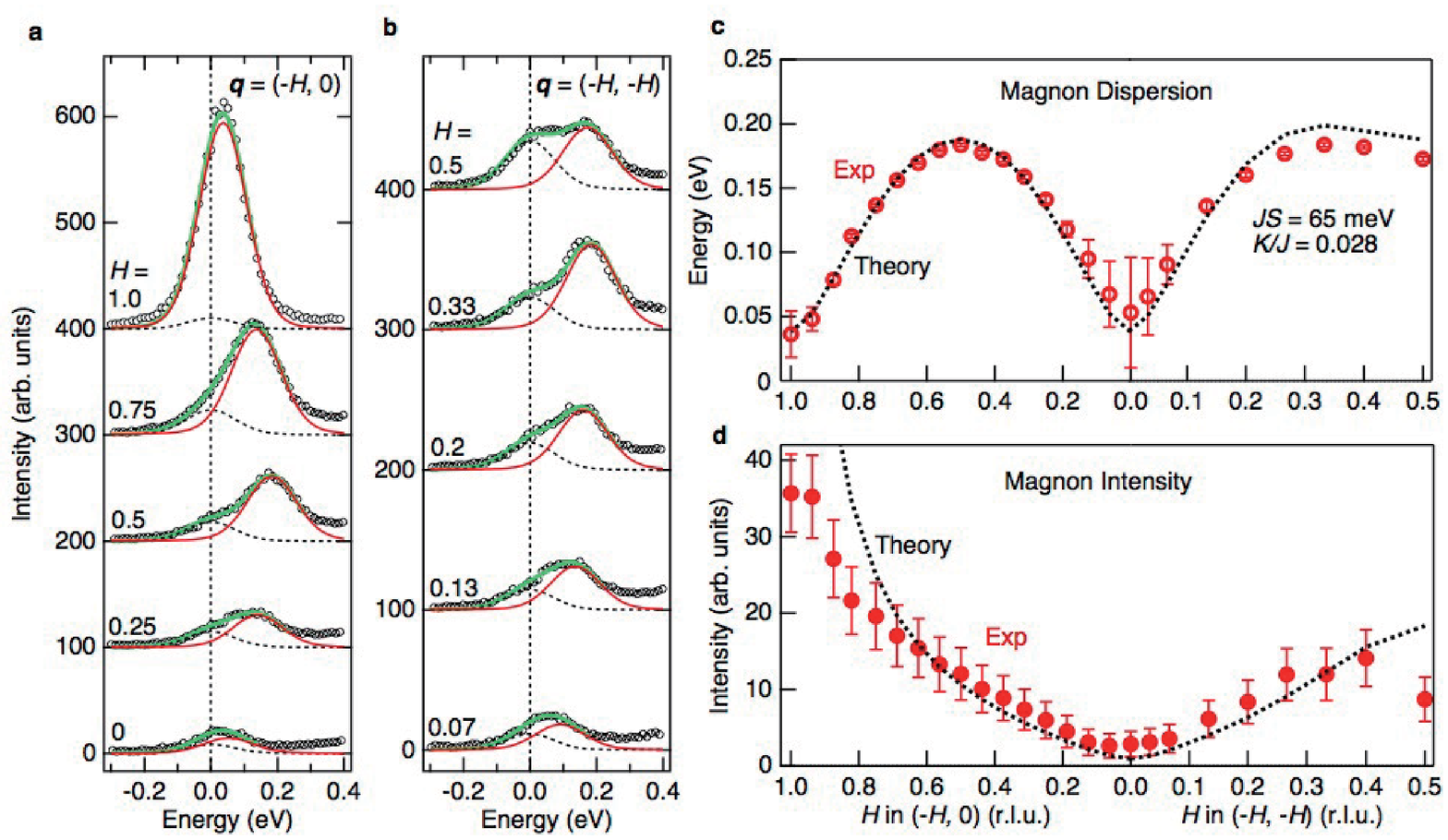}
  \caption*{\raggedright\textbf{Figure 3 $\vert$ Magnon dispersion and intensity}. {\bf a,b}, Decomposition of low-energy IRIXS spectra into the elastic background (dotted black line) and magnon peak (red line). The spectra were fitted to two Lorentzian functions convoluted with the experimental resolution of 140 meV. The total fitted spectra are shown as green lines. {\bf c}, Magnon dispersion as a function of in-plane ${\bm q}$. The error bars represent the standard deviation of the fitting procedure. The dotted line shows the results of a fit of the magnon dispersion to the spin-wave theory defined in equation (1). {\bf d}, Magnon intensity. The dotted line indicates the RIXS intensity calculated from the spin-wave theory with the best-fit parameters. }  \label{Fig2}
 \end{figure}

To determine the exchange interactions between Ru spins from the experimental magnon dispersion, we use the following minimal spin Hamiltonian:
\begin{equation}
{\cal H}=J\sum_{<ij>}{\bm S}_{i}\cdot{\bm S}_{j}-K\sum_{i} S_{iz}^{\; 2},
\end{equation}
where $J$ is an isotropic Heisenberg interaction between nearest-neighbor Ru spins and $K$ is a uniaxial single-ion anisotropy term which orients the spins along the $c$-axis ($K>0$). The inter-plane coupling is assumed to be weak and is neglected. We apply the linear spin-wave approximation to the Hamiltonian to obtain the theoretical magnon dispersion (see Supplementary Information). There are two fitting parameters, $JS$ and $K/J$, and the best fit is obtained with $JS=65$ meV and $K/J=0.028$ (dotted line in Fig. \arabic{Fig3}c). The excellent agreement between the fit results and the IRIXS data demonstrates that the two-dimensional nearest-neighbor Heisenberg interaction and the small anisotropy term of $KS=1.8$ meV are sufficient to describe the magnetic excitations. The slight discrepancy between experimental and theoretical curves along the (-$H$, -$H$) direction can be improved by introducing a second-nearest-neighbor coupling $J_{2}$, but it is clear that $J_{2}$ is much smaller than $J$. These $JS$ and $K/J$ values are reproduced by density-functional LDA+$U_{\text{eff}}$ calculations with moderate $U_{\text{eff}}=1.5$ eV (see Supplementary Information), supporting the model used and the fitting procedure. The same calculations also reproduce the electron-hole continuum revealed in the IRIXS spectra at high energies.


Figure \arabic{Fig3}d shows a comparison between the experimental magnon intensity (defined as the total area of the magnon peak) and the intensity calculated from RIXS operators for $t_{2g}$ orbital systems \cite{Kim.B_etal.Phys.-Rev.-B2017} (see Supplementary Information for details). The ${\bm q}$ dependence for small $\vert H\vert$ is well reproduced by the theory, and the deviations at larger $\vert H\vert$ can be attributed to experimental uncertainties arising from the comparable dimensions of the IRIXS beam spot and the single-crystal samples. 
For small $\theta$ angles (large $|H|$), the small crystal captures only a part of the incident photons, so that the scattering intensity is reduced. 



Finally, we discuss the origin of the pronounced peaks at $\sim$ 0.65 eV and the mechanism of moment reduction in SrRu$_{2}$O$_{6}$. In the $t^{\; 3}_{2g}$ electronic configuration in the cubic crystal field, the intra-ionic Coulomb interaction (intra-orbital $U$ and inter-orbital $U^{\prime}$) and the Hund's rule coupling $J_{H}$ produce several multiplets. The ground state is given by the orbitally-nondegenerate $^{4}A_{2g}$ state with high-spin $S=3/2$. The low-spin ($S=1/2$) multiplets $^{2}E_{g}$ and $^{2}T_{1g}$ are located at the excitation energy $3J_{H}$ (where the relation $U^{\prime}=U-2J_{H}$ is assumed), and the $^{2}T_{2g}$ multiplet at $5J_{H}$. We thus assign the 0.65 eV peaks to a spin-state transition from the $^{4}A_{2g}$ to the $3J_{H}$ multiplets (see Fig. \arabic{Fig2}d), based on the strong intensity and the absence of energy dispersion. We thus obtain an experimental value of $J_{H}\sim 0.2$ eV, which is significantly smaller than the Ru$^{5+}$ ionic value. This indicates strong hybridization between the Ru $t_{2g}$ and O $2p$ orbitals, reducing the magnetic moment from the value expected for isolated $S=3/2$ ions.


The transition to the high-energy $5J_{H}$ multiplet ($> 1$ eV) is not clearly resolved in the IRIXS data. In order to explain this observation, we consider the trigonal distortion $\Delta$ (see Supplementary Information).  Positive $\Delta$ (trigonal compression) enhances the intensity of the $3J_{H}$ transitions, whereas it strongly reduces that of the $5J_{H}$ transitions, so that it is hidden in the electron-hole continuum. Importantly, positive $\Delta\sim 0.25$ eV naturally reproduces the single-ion anisotropy $K$ [the second term in eq. (1)] that opens the magnon gap. Since the orbital angular momentum $L=0$ in the $S=3/2$ ground state, first-order perturbation by the spin-orbit coupling does not lift its fourfold degeneracy. However, second-order perturbations involving higher-energy multiplet levels split the ground state quartet into a lower-lying $S_{z}=\pm 3/2$ doublet and higher-energy $S_{z}=\pm 1/2$ states, stabilizing the $c$-axis spin orientation.



In conclusion, we have successfully observed dispersive magnons in SrRu$_{2}$O$_{6}$ with Ru $L_{3}$ RIXS. The magnon dispersion is well described by the spin-wave theory of the two-dimensional Heisenberg Hamiltonian with a small anisotropy term. The predominance of nearest-neighbor exchange interactions obviates the need to invoke extended molecular orbitals which are expected to generate longer-range interactions \cite{Streltsov.S_etal.Phys.-Rev.-B2015}. On top of the electron-hole continuum above the Mott gap, we observe well-defined non-dispersive spectral features at $\sim$ 0.65 eV, which originate from local spin-state transitions between $S=3/2$ to $S=1/2$ configurations. From an analysis of these data we estimate $J_{H} \sim$ 0.2 eV, which is significantly smaller than the Ru$^{5+}$ ionic value. This indicates strong hybridization between the Ru $t_{2g}$ and O $2p$ orbitals and explains the reduced ordered moment in SrRu$_{2}$O$_{6}$, in agreement with theoretical work \cite{Streltsov.S_etal.Phys.-Rev.-B2015,Singh.D_etal.Phys.-Rev.-B2015,Hariki.A_etal.Phys.-Rev.-B2017,Okamoto.S_etal.Sci.-Rep.2017}. More broadly, our direct observation of Hund's-rule multiplets and the excellent description of magnons based on a Hamiltonian with only nearest-neighbor exchange interactions imply that SrRu$_{2}$O$_{6}$ is in the local-moment regime, albeit close to a metal-insulator transition that explains its extraordinary high N\'eel temperature. The results demonstrate the potential of the newly developed IRIXS methodology to deepen our understanding of exotic magnetism in $4d$ transition metal compounds.

\section*{Methods}

Single crystals of SrRu$_{2}$O$_{6}$ were grown by a hydrothermal method. Sr(OH)$_{2}$$\cdot$8H$_{2}$O (99\%, Alfa Aesar, 0.130 g), KRuO$_{4}$ (97\%, Alfa Aesar, 0.200 g) and 10 ml of deionized water were sealed in a Teflon-lined stainless steel autoclave with inner volume of 23 ml (Parr Instrument Company). The autoclave was heated to 473 K for 48 h and then cooled naturally to room temperature. Before the IRIXS experiments, the single crystals were aligned using an in-house x-ray diffractometer. The photograph of a SrRu$_{2}$O$_{6}$ single crystal (Fig. \arabic{Fig1}c) was taken using a camera attached to a Jobin Yvon LabRam HR800 micro-Raman spectrometer.

The RIXS experiments were performed at the newly-built IRIXS instrument at the P01 beamline of PETRA III at DESY, which delivers a high flux of intermediate-energy photons. The incident energy was tuned to the Ru $L_{3}$ absorption edge (2837.5 eV) and monochromatized using an asymmetrical Si(111) channel-cut  ($+$,$-$) monochromator ($\Delta E$ = 100 meV). The polarization of the incident beam was in the horizontal scattering plane ($\pi$ polarization). The polarization of the scattered x-rays was not analyzed. The x-rays were focused to a beam spot of 20 $\times$ 20 $\mu$m$^{2}$. Scattered photons from the sample were collected at 2$\theta$ $=$ 90$^{\circ}$ (horizontal scattering geometry) using a SiO$_{2}$ (10$\bar{2}$) ($\Delta E$ = 60 meV) diced spherical analyzer with a 1 m arm, equipped with a rectangular (100 (H) $\times$ 36 (V) mm$^{2}$) mask and a CCD camera, both placed in the Rowland geometry.  The exact position of the elastic (zero energy loss) line was determined by measuring non-resonant scattering spectra from silver paint deposited next to the samples. The overall energy resolution of the IRIXS spectrometer at the Ru $L_{3}$-edge was 140 meV, as determined from the full-width-half-maximum of the silver paint spectra. 
The measurements were performed at 10 K, well below $T_{N}$.


\section*{Acknowledgements}
We thank I. I. Mazin for stimulating discussions. The project was supported by the European Research Council under Advanced Grant No. 669550 (Com4Com). We acknowledge DESY (Hamburg, Germany), a member of the Helmholtz Association HGF, for the provision of experimental facilities. The experiments were carried out at the beamlines P01 and P09 of PETRA III at DESY. H.S. and K.U. acknowledge financial support from the JSPS Research Fellowship for Research Abroad. H.S. is partially supported by the Alexander von Humboldt Foundation. 

\section*{Author contributions}
H.S., H.G., K.U., Z.Y., M.M. and H.Y. performed the RIXS experiments. H.I., J.N. and H.T. grew SrRu$_{2}$O$_{6}$ single crystals and performed sample characterization. H.G., H.C.W. and H.Y. designed the beamline and IRIXS spectrometer. H.S., H.G. and K.U. performed the magnetic REXS experiment with the help of J.A.S. and S.F.. H.S. analyzed the experimental data. H.L., H.K., D.K., A.Y., B.J.K. and G.K. carried out the theoretical calculations and contributed to the interpretation of the experimental data.  H.S. and B.K. wrote the manuscript with comments from all co-authors. B.K. initiated and supervised the project.

\section*{Competing interests}
The authors declare no competing interests.

\section*{Data availability}
The data sets generated during and/or analysed during the current study are available from the corresponding author on reasonable request.

\bibliography{SRO_RIXS}

\end{document}